\documentclass[letter,onecolumn]{jpsj3}%

\usepackage{amsfonts}
\usepackage{amsmath}
\usepackage{amssymb}
\usepackage{graphicx}%
\begin{document}

\title{Ferromagnetic instability and finite-temperature properties of two-dimensional
electron systems with van Hove singularities}
\author{A. A. Katanin$^{1}$, H. Yamase$^{2}$, and V. Yu. Irkhin$^{1}$ }

\inst{$^{1}$Institute of Metal Physics, 620041 Ekaterinburg, Russia \\
$^{2}$National Institute for Materials Science, Tsukuba 305-0047, Japan}

\abst{We study a ferromagnetic tendency in the two-dimensional Hubbard model
near van Hove filling by using a functional renormalization-group method.
We compute temperature dependences of magnetic susceptibilities
including incommensurate magnetism. The ferromagnetic tendency
is found to occur in a dome-shaped region around van Hove filling
with an asymmetric property:
incommensurate magnetism is favored near the edge of the dome
above van Hove filling whereas a first-order-like transition to the ferromagnetic
ground state is expected below van Hove filling.
The dome-shaped phase diagram is well captured in the Stoner theory
by invoking a smaller Coulomb interaction.
Triplet $p$-wave superconductivity tends to develop at low temperatures
inside the dome and extends more than the ferromagnetic region above van Hove filling.}

\kword{ferromagnetism, renormalization group, Stoner theory,
incommensurate magnetism, triplet superconductivity, Hubbard model}

\maketitle

The stability of itinerant-electron ferromagnetism is an important problem of
correlated $d$-electron systems \cite{Vons,Irk2}. Stoner proposed a simple
approach to treat the ferromagnetic instability, introducing an averaged
magnetic field of electrons at each lattice site. His theory is, however,
insufficient to describe adequately even magnets with small magnetic moments,
especially at finite temperatures.
The transition temperatures appear too high in comparison with experimental
data and the temperature dependence of the magnetic susceptibility is not
described correctly.

Dzyaloshinskii, Kondratenko \cite{DK} and Moriya \cite{Moriya} proposed the
spin-fluctuation theory considering a contribution of paramagnons to
thermodynamic properties. They improved the Stoner theory, especially for
finite temperature properties of weak and nearly ferromagnetic systems. The
condition for ferromagnetic order in the ground state is also corrected by
their theory. In the regime of strong Coulomb repulsion the stability of
ferromagnetism was considered in the pioneering studies by Nagaoka
\cite{Nagaoka} and Roth \cite{Roth}, which showed the existence of a saturated
ferromagnetic region close to half filling and non-saturated ferromagnetism
further away from it. The region of
the saturated ferromagnetism was investigated in more details within various
approximations in numerous works (see discussion and references in Ref.
\cite{Review}).

Two-dimensional systems offer an interesting aspect to study the stability
condition of a ferromagnetic ground state and its finite-temperature
properties. Because of van Hove singularities in the density of states
ferromagnetism
can be realized within a one-band model even for moderately small Coulomb
repulsion $U$ \cite{VH-ThreeD}. This allows to apply perturbative techniques
to this problem to verify and understand results of other approaches. At
finite temperatures, the continuous symmetry breaking is absent in two
dimensions due to
fluctuation effects, resulting in fulfillment of the Mermin-Wagner theorem;
the Curie-Weiss law is expected to be accordingly modified.

Another peculiarity of two-dimensional systems is that the ferromagnetic
order
in two dimensions can be destabilized not only by quantum magnetic
fluctuations, \text{e.g.}, via spin-wave excitations, which are nearly
commensurate, but also by incommensurate fluctuations. The importance of
incommensurate magnetic fluctuations at weak and moderate Coulomb interaction
was recently emphasized within the quasistatic approach \cite{QS,Our},
mean-field theory \cite{MFQ}, and a renormalization-group approach
\cite{RG_Ferro}. These approaches showed that in a large part of the phase
diagram spanned by the electron density $n$ and Coulomb interaction $U$ the
ferromagnetic order is replaced by an incommensurate one.

Therefore, it is important to consider systematically the effect of
commensurate and incommensurate magnetic fluctuations on the ground-state
phase diagram and finite temperature properties of ferromagnetism in two
dimensions. In an attempt to investigate this problem, we use a functional
renormalization-group (fRG) approach in the \textit{symmetric phase}
\cite{Metzner1,SalmHon,SalmHon1,KK} and study temperature dependences of
magnetic susceptibilities. By extrapolating these results to the limit of zero
temperature, one can obtain expected properties such as possible instabilities
of the ground state. This procedure was recently applied to studying a
possibility of antiferromagnetic and superconducting instabilities near half
filling \cite{Katanin}.


\textit{Model and method}. We consider the two-dimensional (2D) $t$%
-$t^{\prime}$ Hubbard model $H_{\mu}=H-(\mu-4t^{\prime})N$ with
\begin{equation}
H=-\sum_{ij\sigma}t_{ij}c_{i\sigma}^{\dagger}c_{j\sigma}+U\sum_{i}%
n_{i\uparrow}n_{i\downarrow}\,,\label{H}%
\end{equation}
where $t_{ij}=t$ for nearest neighbor (nn) sites $i$, $j$, and $t_{ij}%
=-t^{\prime}$ for next-nn sites ($t,t^{\prime}>0$) on a square lattice;
$c_{i\sigma}^{\dagger}(c_{i\sigma})$ creates (annihilates) an electron with
spin $\sigma$ at $i$ site; $n_{i\sigma}=c_{i\sigma}^{\dagger}c_{i\sigma}$ and
$N=\sum_{i\sigma}n_{i\sigma}$. For convenience we have shifted the chemical
potential $\mu$ by $4t^{\prime}$.
We choose $t^{\prime}/t=0.45$ as a typical value, for which the ferromagnetic
instability is favored \cite{SalmHon1,KK,Salmhofer}. We employ the fRG
approach for the one-particle irreducible generating functional and choose the
temperature $T$ as a natural cutoff parameter as proposed in Ref.
\cite{SalmHon1}. Neglecting the frequency dependence of interaction vertices,
the RG differential equation for the interaction vertex $V_{T}\equiv$
$V(\mathbf{k}_{1},\mathbf{k}_{2},\mathbf{k}_{3},\mathbf{k}_{4})$ (with the
momenta $\mathbf{k}_{i}\,$, which are supposed to fulfill the conservation law
$\mathbf{k}_{1}+\mathbf{k}_{2}=\mathbf{k}_{3}+\mathbf{k}_{4}$) has the form
\cite{SalmHon1,KK,Katanin}
\begin{equation}
\frac{\mathrm{d}V_{T}}{\mathrm{d}T}=-V_{T}\circ\frac{\mathrm{d}L_{\mathrm{pp}%
}}{\mathrm{d}T}\circ V_{T}+V_{T}\circ\frac{\mathrm{d}L_{\mathrm{ph}}%
}{\mathrm{d}T}\circ V_{T}\,,\label{dV}%
\end{equation}
where $\circ$ is a short notation for summations over intermediate momenta and
spins, $L_{\text{ph,pp}}$ stand for particle-hole and particle-particle
bubbles. The ferromagnetic (FM), triplet $p$-wave superconducting ($p$SC) and
incommensurate magnetic (\textbf{Q}) susceptibilities can be calculated as%
\begin{align}
&  \displaystyle{\frac{\mathrm{d}\chi_{m}}{\mathrm{d}T}}%
\begin{array}
[c]{c}%
=
\end{array}
\sum_{\mathbf{k}^{\prime}}\mathcal{R}_{\mathbf{k}^{\prime},\pm\mathbf{k}%
^{\prime}+\mathbf{q}_{m}}^{m}\mathcal{R}_{\mathbf{k}^{\prime},\pm
\mathbf{k}^{\prime}+\mathbf{q}_{m}}^{m}\displaystyle{\frac{\mathrm{d}%
L_{\text{ph,pp}}(\mathbf{k}^{\prime};\mathbf{q}_{m})}{\mathrm{d}T}}%
,\label{dH}\\
&  \displaystyle{\frac{\mathrm{d}\mathcal{R}_{\mathbf{k},\pm\mathbf{k-q}_{m}%
}^{m}}{\mathrm{d}T}}%
\begin{array}
[c]{c}%
= \mp
\end{array}
\sum_{\mathbf{k}^{\prime}}\mathcal{R}_{\mathbf{k}^{\prime},\pm\mathbf{k}%
^{\prime}+\mathbf{q}_{m}}^{m}\Gamma_{m}^{T}(\mathbf{k},\mathbf{k}^{\prime
})\displaystyle{\frac{\mathrm{d}L_{\text{ph,pp}}(\mathbf{k}^{\prime
};\mathbf{q}_{m})}{\mathrm{d}T}},\nonumber
\end{align}
where the three-point vertices $\mathcal{R}_{\mathbf{k,k^{\prime}}}^{m}$
describe the propagation of an electron in a static external field, $m$
denotes a type of instability, $\mathbf{q}_{\text{FM,\textit{p}SC}}=0$
and $\mathbf{q}_{\mathbf{Q}}=\mathbf{Q;}$ upper signs and ph correspond to the
magnetic instabilities, lower signs and pp to the superconducting instability;%
\begin{equation}
\Gamma_{m}^{T}(\mathbf{k},\mathbf{k}^{\prime})=\left\{
\begin{array}
[c]{cl}%
V(\mathbf{k},\mathbf{k}^{\prime},\mathbf{k}^{\prime}+\mathbf{q}_{m}%
,\mathbf{k}-\mathbf{q}_{m}) & m=\text{FM or \textbf{Q},}\\
V(\mathbf{k},-\mathbf{k},\mathbf{k}^{\prime},-\mathbf{k}^{\prime}) &
m=p\text{SC.}%
\end{array}
\right.
\end{equation}

Eqs. (\ref{dV}) and (\ref{dH}) are solved with the initial conditions
$V_{T_{0}}(\mathbf{k}_{1},\mathbf{k}_{2},\mathbf{k}_{3},\mathbf{k}_{4})=U$,
$\mathcal{R}_{\mathbf{k,k+q}_{m}}^{m}=f_{\mathbf{k}}$ and $\chi_{m}=0$; the
initial temperature is chosen as large as $T_{0}=10^{3}t$. The function
$f_{\mathbf{k}}$ belongs to one of the irreducible representations of the
point group of the square lattice, \text{i.e.}, $f_{\mathbf{k}}=1$ for the
magnetic instabilities and $f_{\mathbf{k}}=\sin k_{x,y}/A$ for the $p$SC, with
$A$ being a normalization coefficient.
We also discretize the momentum space in $N_{p}=48$ patches using the same
patching scheme as in Ref. \cite{SalmHon1}. This reduces the
integro-differential equations (\ref{dV}) and (\ref{dH}) to a set of 5824
differential equations, which were solved numerically. In the present paper we
perform the renormalization-group analysis down to the temperature
$T_{\text{RG}}^{\min},$ where vertices reach a maximal value; we choose
$V_{\max}=18t$.

To characterize the strength of fluctuations of the order parameter $m$, we
introduce the temperature $T_{m}^{\ast}$ which is defined by the condition
$\overline{\chi}_{m}^{-1}(T_{m}^{\ast})=0.$ Here $\overline{\chi}_{m}^{-1}(T)$
is the inverse susceptibility of the order parameter $m$, analytically
extrapolated to the region $T<T_{\text{RG}}^{\min}$ (see Ref. \cite{Katanin}
for details). We interpret $T_{m}^{\ast}$ as a crossover temperature to the
regime of strong correlations rather than a phase transition temperature,
since the finite temperature transition is prohibited by the Mermin-Wagner
theorem, inducing non-analytical corrections to susceptibilities,
\text{e.g.}, 
$\chi_{\text{FM,\textbf{Q}}}^{-1}(T)\propto e^{-A_{\text{FM,\textbf{Q}}}/T}$ at low
$T$ as a consequence of an exponentially large correlation length (see,
\text{e.g.}, Ref. \cite{Tremblay}).

To compare the obtained results with the Stoner theory we also study
Hamiltonian (\ref{H}) in the mean-field approximation, by decoupling
$n_{i\uparrow}n_{i\downarrow}\rightarrow\langle n_{\uparrow}\rangle
n_{i\downarrow}+n_{i\uparrow}\langle n_{\downarrow}\rangle-\langle
n_{\uparrow}\rangle\langle n_{\downarrow}\rangle$. The second-order transition
temperature (Curie temperature) is obtained by the condition $\chi_{\text{FM}%
}^{-1}(T_{C}^{\mathrm{MF}})=0$. To search for possible first-order
transitions, we also compute the Landau free energy as a function of
magnetization for fixed chemical potentials.


\textit{Results.} In Fig.~1(a) we present the results of the fRG approach for
the inverse magnetic susceptibilities, obtained at $U=4t$ and different
fillings above van Hove filling $n=0.465$ ($\mu=0$). Above the crossover
temperature the susceptibilities follow approximately the Curie-Weiss law with
Curie temperatures replaced by $T_{\text{FM}}^{\ast}$. With increasing $\mu$,
the inverse susceptibilities increase, indicating that ferromagnetic
fluctuations are weakened.
Consequently, the crossover temperature $T_{\text{FM}}^{\ast}$ decreases and
vanishes at $\mu\simeq0.052t$, implying the quantum phase transition from the
ferromagnetic to paramagnetic phase. This transition is, however, likely
preempted by entering into the phase with strong incommensurate fluctuations,
since we confirmed by calculations of $\overline{\chi}_{\mathbf{Q}},$ which
yield results similar to Fig. 1(a), that in the range $\mu\in(0.05,0.055)t$ a
condition $T_{\mathbf{Q}}^{\ast}>T_{\text{FM}}^{\ast}\ $is fulfilled for some
of the vectors $\mathbf{Q}=(Q,Q)$ with small $Q.$ Therefore, for the ground
state above van Hove filling one can expect two successive quantum phase
transitions: from the ferromagnetic to an incommensurate and then to the
paramagnetic phase with increasing $\mu.$ A direct phase transition from the
ferromagnetic to paramagnetic phase having strong incommensurate magnetic
fluctuations is yet also possible. Further away from van Hove filling we
obtain a non-monotonic temperature dependence of the inverse susceptibility
with a minimum, which is followed by a maximum for $\mu\in(0.065,0.085)t$.
While the minima of the inverse susceptibility away from van Hove filling are
produced by thermal excitations of states near van Hove singularity, an
interpretation of the maxima which possibly yield a `reappearance' of magnetic
order far away from van Hove filling is not so straightforward. We return to
the discussion of this peculiarity below. Even further away from van Hove
filling ($\mu>0.09t$) the flow can be continued down to very low temperatures,
such that $T_{\text{RG}}^{\mathrm{\min}}\approx0.$

\begin{figure*}[tb]
\begin{center}
\includegraphics[width=13cm]{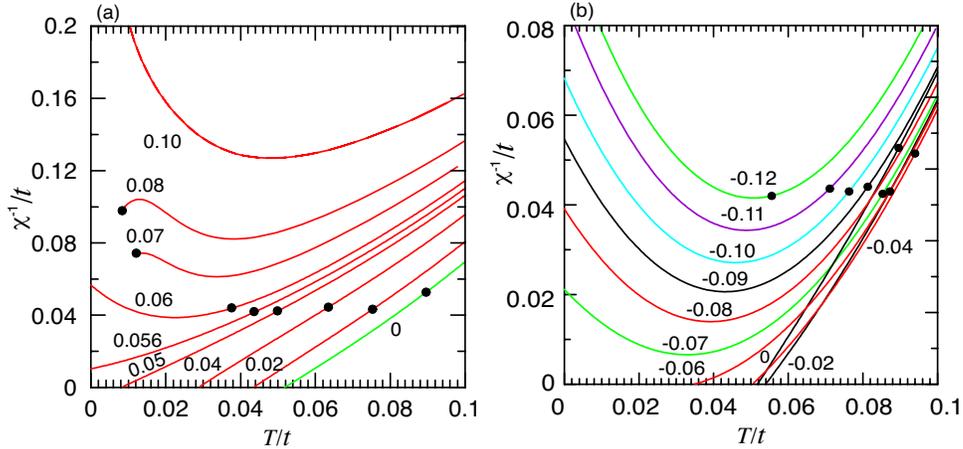}
\end{center}
\caption{(Color online) Temperature dependences of the inverse magnetic
susceptibility at $U=4t$ and different values of the chemical potential
(marked by numbers) for the Fermi level above (a) and below (b) van Hove
energy. Dots denote the temperature $T_{\text{RG}}^{\min}$, at which the fRG
flow is stopped; note that $T_{\mathrm{RG}}^{\mathrm{min}} \approx0$ for
$\mu=0.10t$. Below $T_{\mathrm{RG}}^{\mathrm{min}}$, the susceptibility is
analytically extrapolated. }%
\end{figure*}

Let us consider now the situation below van Hove filling, which is distinct
from the situation above van Hove filling. The obtained temperature dependence
of the inverse susceptibilities [see Fig.~1(b)] shows larger bending than that
for the Fermi level above van Hove filling; the 
inverse susceptibilities show deviations from the Curie-Weiss law already at
$T>T_{\text{FM}}^{\ast}$. 
Moreover, the obtained crossover temperature $T_{\text{FM}}^{\ast}$ sharply
drops to zero at $\mu\simeq-0.065t.$ The precise form of the dependence of
$T_{\text{FM}}^{\ast}(\mu)$ in this region depends on the details of
analytical extrapolation. One can obtain either discontinuous or sharp
continuous transition with vanishing $T_{\text{FM}}^{\ast}$. Since
$T_{\text{FM}}^{\ast}$ should be interpreted as a \textit{crossover}
temperature characterizing magnetic properties of the \textit{ground state} in a 2D
system, this necessarily implies a sharp change of those around
$\mu\approx-0.065t$, in particular a possibility of a \textit{first-order}
phase transition from the ferromagnetic to paramagnetic ground state. Studying
the susceptibilities of incommensurate magnetic orders shows that we always
obtain $T_{\mathbf{Q}}^{\ast}<T_{\text{FM}}^{\ast}$ below van Hove filling,
\text{i.e.}, incommensurate magnetic fluctuations are not expected to change 
our conclusion.

The resulting finite-temperature phase diagram in the $T$-$\mu$ variables is
shown in Fig.~2. $T_{\mathrm{FM}}^{\ast}$ forms a dome-shaped line, which is
asymmetric with respect to the van Hove energy ($\mu=0$). The incommensurate
magnetic tendency occurs near the edge of the dome above the van Hove energy
whereas such a feature is not seen below the van Hove energy and instead a
first-order-like transition to (commensurate) ferromagnetism is expected as a
ground state. In Fig.~2, we also plot $T^{\mathrm{min}}_{\mathrm{RG}}(\mu),$
which is usually interpreted as a crossover temperature to an ordering
tendency\cite{Metzner1,SalmHon,SalmHon1}, by rescaling both temperature and
$\mu$ by a factor of 1.6. The line of $T^{\mathrm{min}}_{\mathrm{RG}}$ almost
coincides with that of $T_{\mathrm{FM}}^{\ast}$, which suggests that our
extrapolation procedure to obtain $T_{\mathrm{FM}}^{\ast}$ is performed in a
reasonable way. Employing the same extrapolation procedure for the
non-monotonic dependence of the inverse susceptibilities seen in Fig.~1(a)
around $\mu\simeq0.07t$, we obtain the `second', tiny, ferromagnetic region as
shown in Fig. 2. We are however not aware of the physical explanation of 
this possible `reappearance' of ferromagnetic order, and leave it for future studies. 

Since triplet superconductivity is expected around a ferromagnetic
state\cite{SalmHon1}, we also compute $\chi_{p\mathrm{SC}}$ as a function of
$T$ in the same fashion of Fig.~1. In Fig. 2 we plot $T_{\text{$p$SC}}^{\ast}%
$, which, similar to previous results \cite{SalmHon1}, is smaller than
$T_{\text{FM}}^{\ast}$ around van Hove filling and survives even far above van
Hove filling, where $p$-wave superconductivity is therefore expected in the
ground state. On the other hand, we do not find an appreciable $T_{\text{$p$%
SC}}^{\ast}$ away from the ferromagnetic phase below van Hove filling. In the
region $\mu\in(-0.07,0.05)t$ the coexistence of ferromagnetism and $p$-wave
superconductivity is possible, which is a subject of future studies.

\begin{figure}[tbh]
\begin{center}
\includegraphics[width=7.5cm]{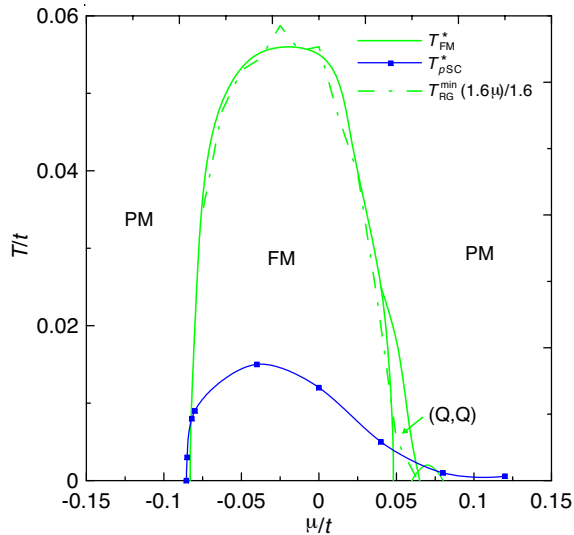}
\end{center}
\caption{(Color online) Phase diagram obtained in the fRG study for $U=4t$ in
the plane of $\mu$ and $T$. PM denotes the paramagnetic phase.}%
\end{figure}

We now compare the fRG results to those obtained in the Stoner theory, which
considers only a (commensurate) ferromagnetic instability. Since the Stoner
theory predicts much higher transition temperatures $T_{C}^{\text{MF}}$ and a
much broader concentration range of ferromagnetism, we consider smaller
$U=U^{\text{eff}},$ which is chosen such that $\max_{\mu}T_{C}^{\text{MF}}%
(\mu;U^{\text{eff}})=\max_{\mu}T_{\text{FM}}^{\text{*}}(\mu;U)$. For $U=4t$ we
obtain $U^{\text{eff}}\simeq1.7t$. This renormalization of $U$ is due to
fluctuations, mainly caused by particle-particle scattering processes,
similar to the original idea by Kanamori\cite{Kanamori}.
In Fig.~3, we see that not only the height, but also the position and the
width of the ferromagnetic region are in good agreement with those in the fRG
approach. Furthermore, below van Hove filling, the possibility of the
first-order-phase transition as a function of $\mu$ (Fig.~2) within the fRG is
also in agreement with the results of `renormalized' Stoner theory, since such
a transition is transformed to a phase separation in terms of electronic
density \cite{Yamase,Our}. Above van Hove filling the renormalized Stoner
theory yields a region of phase separation, while the fRG predicts the strong
tendency of incommensurate magnetism in the corresponding region. Except for
this difference, it is remarkable that the ferromagnetic tendency obtained in
the fRG is captured very well in the renormalized Stoner theory with
$U^{\text{eff}}$, which is $n$ and $T$-independent.

\begin{figure}[tb]
\begin{center}
\includegraphics[width=7cm]{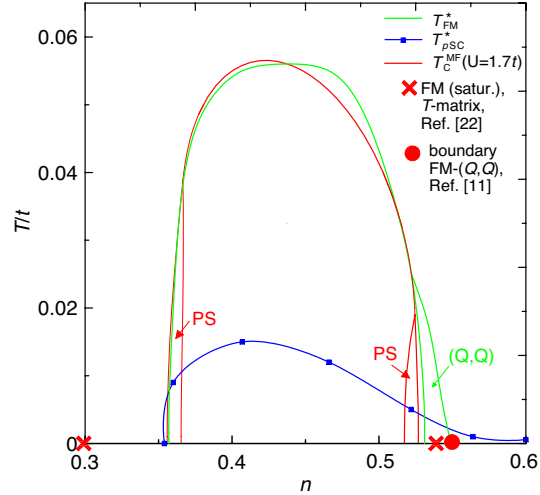}
\end{center}
\caption{(Color online) Comparison of the fRG phase diagram in the $n$-$T$
coordinates with the Stoner theory and with the ground-state results of
mean-field approach including incommensurate magnetic orders \cite{MFQ} and
$T$-matrix approximation \cite{Tmatrix}. PS denotes a region of phase separation.}%
\end{figure}

Considering incommensurate magnetic order in the mean-field theory yields 
a variety of the ground states in the plane of $n$ and $U$, as revealed
in Ref.~\cite{MFQ}. In particular, the mean-field theory \cite{MFQ}
already predicts incommensurate magnetism above van Hove filling and
(commensurate) ferromagnetism roughly below it, for $U=4t$ the ground state
changes
at $n=0.55$ through a second-order transition.
Naturally, the mean-field treatment of Ref.~\cite{MFQ} predicts magnetic order
in a very broad region ($0<n<0.9$), which is obviously an overestimation,
typical for a mean-field theory. In particular,
a broad incommensurate region ($0.55<n<0.9$) is replaced by a very narrow
region with strong incommensurate fluctuations in the fRG approach.
Finally we find that the positions of the quantum phase transitions obtained
in the fRG approach show reasonable agreement with the results of $T$-matrix
approach \cite{Tmatrix}. The $T$-matrix approach, however, misses the
importance of incommensurate magnetic fluctuations above van Hove filling.


\textit{Conclusion.} We have studied finite temperature phase diagrams and
possible ground-state properties of the one-band Hubbard model at different
chemical potentials or fillings. The results are compared to those in the
Stoner theory, including a possibility of incommensurate magnetic order. We
have found that a ferromagnetic tendency occurs in a dome-shaped region around
van Hove filling in an asymmetric way: incommensurate magnetism is favored
near the edge of the dome above van Hove filling, whereas a first-order-like
transition to a (commensurate) ferromagnetic ground state is expected below
van Hove filling. The verification of the first-order transition however
requires further development of the present fRG approach, \text{e.g.}, its
extension to the symmetry-broken phase. At finite temperatures we observe
deviations from the Curie-Weiss behaviour of susceptibilities, which are
mostly pronounced below van Hove filling and seen already above the
temperature of the crossover to the regime of strong magnetic correlations. In
agreement with previous studies, a ferromagnetic tendency is accompanied by
development of triplet superconductivity at low temperatures.
While our study indicates a pure $p$-wave superconducting state far above van
Hove filling, we cannot address whether the ground state is the coexistence of
ferromagnetism and triplet superconductivity in a region where a 
ferromagnetic tendency also occurs. Given that the coexistence is actually
observed for various U-based compounds such as UGe$_{2}$ \cite{saxena00},
URhGe \cite{aoki01} and UCoGe \cite{huy07}, it is an interesting subject to
elucidate a possible coexistence of ferromagnetism and $p$-wave
superconductivity in the ground state of the Hubbard model near van Hove filling.

\textit{Acknowledgements.} The authors thank J. Bauer for valuable comments.


\begin{thebibliography}{99}                                                                                               %


\bibitem {Vons}S.V. Vonsovski: Magnetism (Wiley, New York, 1974).

\bibitem {Irk2}V.Yu. Irkhin and Yu.P. Irkhin: Electronic structure,
correlation effects and properties of $d$- and $f$-metals and their compounds
(Cambridge International Science Publishing, 2007).

\bibitem {DK}I. E. Dzyaloshinskii and P. S. Kondratenko: Sov. Phys. JETP
\textbf{43} (1976) 1036.

\bibitem {Moriya}T. Moriya: Spin Fluctuations in Itinerant Electron Magnetism
(Springer-Verlag, Berlin, 1985).

\bibitem {Nagaoka}Y. Nagaoka: Phys. Rev. \textbf{147} (1966) 392.

\bibitem {Roth}L. M. Roth: Phys. Rev. \textbf{184} (1969) 451; \textbf{186}
(1969) 428.

\bibitem {Review}M.I. Katsnelson et al., Rev. Mod. Phys. \textbf{80} (2008) 315.

\bibitem {VH-ThreeD}This also applies to the three-dimensional lattices with
the lines of van Hove singularities, leading to divergent density of states,
see S. V. Vonsovskii, M. I. Katsnelson, and A. V. Trefilov: Fiz. Metallov.
Metalloved. \textbf{76} [3] (1993) 4; [4] (1993) 3. We do not however consider
this situation in the present paper, since, as a rule, it requires also
treatment of multi-band effects.


\bibitem {QS}P. A. Igoshev, A. A. Katanin, and V. Yu. Irkhin: Sov. Phys. JETP
\textbf{105} (2007) 1043.

\bibitem {Our}P. A. Igoshev, A. A. Katanin, H. Yamase, and V. Yu. Irkhin: J.
Magn. Magn. Mater. \textbf{321} (2009) 899.

\bibitem {MFQ}P. A. Igoshev {\it et al.}: Phys. Rev. B \textbf{81} (2010) 094407.

\bibitem {RG_Ferro}P. A. Igoshev, V. Irkhin, and A. A. Katanin: arXiv: 1012.0125.

\bibitem {Metzner1}C. J. Halboth and W. Metzner: Phys. Rev. B \textbf{61}
(2000) 7364.

\bibitem {SalmHon}C. Honerkamp, M. Salmhofer, N. Furukawa, and T. M. Rice:
Phys. Rev. B \textbf{63} (2001) 035109.

\bibitem {SalmHon1}C. Honerkamp and M. Salmhofer: Phys. Rev. Lett. \textbf{87}
(2001) 187004; Phys. Rev. B\textbf{\ 64} (2001) 184516.

\bibitem {KK}A. A. Katanin and A. P. Kampf: Phys. Rev. B \textbf{68} (2003) 195101.

\bibitem {Katanin}A. A. Katanin: Phys. Rev. B \textbf{81} (2010) 165118.

\bibitem {Salmhofer}C. Husemann and M. Salmhofer: Phys. Rev. B \textbf{79}
(2009) 195125.

\bibitem {Tremblay}Y. M. Vilk and A.-M.S. Tremblay: J. Phys. I (France)
\textbf{7} (1997) 1309.

\bibitem {Kanamori}T. Kanamori: Prog. Theor. Phys. \textbf{30} (1963) 275.

\bibitem {Yamase}H. Yamase, V. Oganesyan, and W. Metzner: Phys. Rev. B
\textbf{72} (2005) 35114.

\bibitem {Tmatrix}R. Hlubina: Phys. Rev. B \textbf{59} (1999) 9600.

\bibitem {saxena00}
S. S. Saxena \textit{et al.}: Nature \textbf{406} (2000) 587.

\bibitem {aoki01}D. Aoki \textit{et al.}: Nature \textbf{413} (2001) 613.

\bibitem {huy07}N. T. Huy \textit{et al.}: Phys. Rev. Lett. \textbf{99} (2007) 067007.
\end{thebibliography}
\end{document}